# The Abundance of Helium in the Source Plasma of Solar Energetic Particles

**Donald V. Reames**

Institute for Physical Science and Technology, University of Maryland, College Park, MD 20742-2431 USA, email: dvreames@umd.edu

**Abstract** Studies of patterns of abundance enhancements of elements, relative to solar-coronal abundances, in large solar energetic-particle (SEP) events, and of their power-law dependence on the mass-to-charge ratio $A/Q$ of the ions, have been used to determine the effective source-plasma temperature $T$ that defines the $Q$-values of the ions. We find that a single assumed value for the coronal reference He/O ratio in all SEP events is often inconsistent with the transport-induced power-law trend of the other elements. In fact, the coronal He/O actually varies rather widely from one SEP event to another. In the large Fe-rich SEP events with $T \approx 3$ MK, where shock waves, driven out by coronal mass ejections (CMEs), have reaccelerated residual ions from impulsive suprathermal events that occur earlier in solar active regions, He/O ≈ 90, a ratio similar to that in the slow solar wind, which may also originate from active regions. Ions in the large SEP events with $T < 2$ MK may be accelerated outside active regions, and have values of $40 \leq$ He/O $\leq 60$. Mechanisms that determine coronal abundances, including variations of He/O, are likely to occur near the base of the corona (at ~ 1.1 $R_S$) and thus to affect both SEPs (at ~2 – 3 $R_S$) and the solar wind. Other than He, reference coronal abundances for heavier elements show little temperature dependence or systematic difference between SEP events; He, the element with the highest first ionization potential, is unique. The CME-driven shock waves probe the same regions of space, at ~2 $R_S$ near active regions, which are also likely sources of the slow solar wind, providing complementary information on conditions in those regions.

## 1. Introduction

The relative abundances of the chemical elements and isotopes have been essential in distinguishing energetic-particle sources and in understanding the physical processes of particle acceleration and transport (*e.g.* Reames, 1999). For solar energetic-particle (SEP) events, there has been a long history distinguishing the small "impulsive" events, with 1000-fold enhancements of $^3$He/$^4$He and of heavy elements ($Z > 50$)/O, associated with magnetic reconnection at the Sun in flares and especially the open fields in jets (Kahler, Reames, and Sheeley 2001, Reames 2002), in contrast with the large "gradual" or long-duration events, where elements with abundances similar, on average, to those of the solar corona, are accelerated at shock waves driven out from the Sun by wide, fast coronal mass ejections (CMEs) (see reviews by Gosling, 1993; Lee, 1997; Reames, 1999, 2013, 2015, 2017a; Desai *et al.*, 2004; Mason, 2007; Kahler, 2007; Cliver and Ling, 2007, 2009; Kahler *et al.*, 2012; Wang *et al.*, 2012). This association of gradual events, the most intense and energetic of the SEP events, with CME-driven shock waves has become well established (see Kahler *et al.*, 1984; Mason, Gloeckler, and Hovestadt, 1984; Kahler, 1992, 1994, 2001; Gopalswamy *et al.*, 2004, 2012; Cliver, Kahler, and Reames, 2004; Lee 2005; Rouillard *et al.*, 2011, 2012; Lee, Mewaldt, and Giacalone, 2012; Desai and Giacalone 2016), although, for a time, this association was complicated by the fact that these shocks can also preferentially reaccelerate residual suprathermal ions from previous impulsive SEP events (Mason, Mazur, and Dwyer 1999; Desai *et al.* 2003, 2006; Tylka *et al.*, 2005; Tylka and Lee, 2006) which may abound near solar active regions.

For many years it has been recognized that the element abundances in gradual SEP events, relative to their photospheric abundances, have an underlying dependence on the first ionization potential (FIP) of the element (*e.g.* Webber 1975; Meyer 1985). This "FIP-effect" was understood to represent abundances of the solar corona itself. Low-FIP (<10 eV) elements are ionized in the photosphere while high-FIP elements are neutral atoms. Since the ions (*i.e.* Mg, Si, and Fe) are much more easily convected from the chromosphere to the corona than the neutrals (*i.e.* He, C, O, and Ne), as by the action of Alfvén waves (Laming 2009, 2015), for example, they become ~3 times as abundant as the neutral atoms when they reach the corona, where all atoms soon become multiply ionized at the ~1–3 MK electron temperature.

Meyer (1985) found that the FIP-effect was similar for all SEP events, but a dependence on the mass-to-charge ratio *A/Q* of the ions, depending upon acceleration and transport, varied from event to event. Using newly available SEP ionization-state measurements of *Q*, Breneman and Stone (1985) showed SEP events that clearly increased or decreased as a power law in *A/Q*. As accelerated ions travel out along the magnetic field from the shock source to an observer, they are scattered by magnetic irregularities, such as Alfvén waves (Parker 1963). This scattering depends upon the magnetic rigidity, or momentum per unit charge, of the particle. If we compare different particle species at the same velocity, their differences in rigidity becomes differences in *A/Q*. Thus an abundance ratio such as Fe/O will increase early in an event, causing it to be depleted later, since Fe, with a higher ratio of *A/Q*, scatters less and spreads out faster than O. Solar rotation turns this variation in time to an additional variation with solar longitude (see examples *e.g.* Reames 2013, 2015). In fact, streaming particles amplify Alfvén waves, greatly increasing the effects of transport in the largest events with high proton intensities (see Ng, Reames, and Tylka 2003; Reames and Ng 2010). For most events, however, diffusion theory, that assumes time-invariant scattering mean free paths *λ*, varying as a power of *A/Q*, is adequate to explain the time dependence of abundance ratios (Reames 2016b). These also vary as a power of *A/Q* that is linear in inverse time, *i.e.* $t^{-1}$.

While transport can redistribute energetic ions in space and time, none are created or destroyed (so long as energy changes from adiabatic deceleration are modest). Hence, if Fe/O is enhanced early in an event because Fe scatters less than O, Fe/O will be depleted later where the Fe has preferentially leaked away. Thus, for many years, it has been common to assume that "coronal" abundances, as a reference for SEP studies, may be recaptured by averaging over a large number of SEP events (*e.g.* Reames, 1995, 2014), recombining the enhanced and depleted abundance regions.

The enhancement or depletion of abundances, relative to these "coronal" values depends upon *A/Q* of each species, but the values of *Q* may depend upon the electron temperature of the source plasma, *T*. While the ionization process need not be isother-mal, it may be possible to find an effective value of *T* with associated *Q*-values that give a pattern like that of the observed element enhancements.

## *1.1 Abundance Enhancements and Source Temperatures*

An early attempt to deduce *T* from abundance enhancements was made by Reames, Meyer, and von Rosenvinge (1994). They noted that in impulsive SEP events C, N, and O, were unenhanced, like He, and were probably also fully ionized with $A/Q \approx 2.0$, while Ne, Mg, and Si formed a group that might correspond to a two-electron closed shell with $A/Q \approx 2.33$, while Fe had $A/Q \sim 3.6$. These ionization levels occur in a coronal temperature range of 3–5 MK. Using more accurate modern abundance measurements, which included the elements from $2 \leq Z \leq 82$, Reames, Cliver, and Kahler (2014a, 2014b, 2015) found that nearly all 111 impulsive SEP events studied had source temperatures of 2.5–3.2 MK apparently from coronal active regions. Impulsive SEPs do *not* share the temperatures of solar flares, typically 10–20 MK, where even Ne, Mg, and Si would become fully stripped so their abundances could not be enhanced relative to He, C, or O.

A graph of *A/Q vs. T* for a variety of elements is shown as Figure 1. As *T* changes, ionization states tend to linger at closed electron shells. As temperature decreases below 3 MK, first O, then N, then C, cross from the fully-ionized shell, like He, to the two-electron "Ne-like" shell. During the same temperature excursion, Ar joins Ca, then S, then Si, then Mg each cross from the "Ne-like" to the "Ar-like" shell.

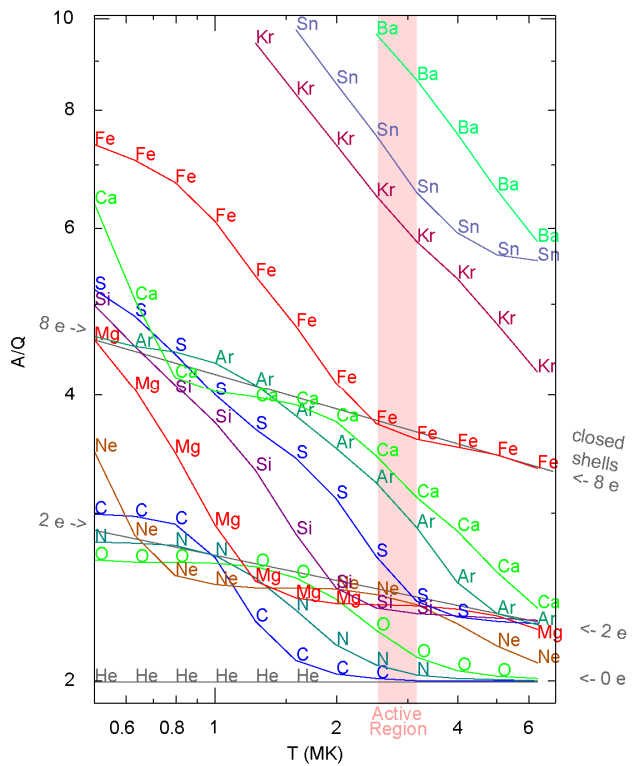

**Figure 1.** *A/Q* varies as a function of the theoretical equilibrium temperature for elements named along each curve. Data are from Mazzotta *et al.* (1998) up to Fe and from Post *et al.* (1977) above. Points are spaced every 0.1 unit of $\log_{10} T$. *A/Q*-values tend to cluster in bands produced by closed electron shells; those with 0, 2, and 8 electrons are shown, He having no electrons. Elements move systematically from one band to another as temperatures change. The shaded pink region corresponds to active-region temperatures found for impulsive SEP events.

If we wish to determine a temperature from observed enhancements, we remember that the pattern of enhancement is a proxy for the pattern of *A/Q*, and ask, is the abundance of C like He (>2 MK), like Ne (<1 MK), or between? Is Si like Ne (>2 MK), or like Ar (<1 MK)? These patterns can uniquely determine the temperature. Procedurally we can least-squares fit log X/O *vs.* log *A/Q* at each temperature on the grid, record the value of $\chi^2$ at each *T*, and select the fit and the value of *T* with the minimum $\chi^2$. Examples of this type of analysis of gradual SEP events are shown and discussed in some detail by Reames (2016a) who found temperatures that were consistent with time throughout 45 gradual SEP events. Reames (2016a) found that 24% of these gradual SEP events had source plasma temperatures of 2.5–3.2 MK, consistent with a strong component of ions originally from impulsive SEP events (jets) from active regions on the Sun, that have been reaccelerated by the shock wave, and 69% with temperatures < 2 MK involving shock acceleration of ambient coronal plasma outside active regions. Nevertheless, all of these SEP events were assumed to have the same underlying "coronal" FIP effect (see Figure 12 and Table 1 of Reames 2014 or Figure 1.6 and Table 1.1 of Reames 2017a).

Using abundance enhancements and their corresponding *A/Q* dependence to determine consistent source temperatures is a relatively new technique that can be applied widely in SEP events. For impulsive SEP events it provides temperatures at the point of acceleration, whereas later direct charge measurements in space may show modification of *Q* when ions pass through material after they leave sources below ~1.5 solar radii (*e.g.* DiFabio *et al.* 2008). For gradual events, mean values of $Q_{Fe}$ near Earth range from 10–14 as determined by direct and geomagnetic measurements (see review by Klecker 2013). These values are consistent with coronal temperatures of ~1.2–2.5 MK. However, measurements of ionization states in SEP events are no longer available near Earth, much less at distant longitudes or for the rarer elements we consider.

## *1.2 What is the "Coronal" Abundance of He?*

The ions from the impulsive events and 2.5–3.2 MK gradual events show evidence of an unusually high ratio of He/C (Reames 2016b). At this temperature He and C should be fully ionized and O, nearly so. So *A/Q* should be ≈2 for all these ions, and the observed ratio should be the source ratio, which is found to correspond to He/O= 91 rather than 57 obtained from comprehensive abundance averages (*e.g.* Reames 1995, 2014). Do these hotter sources have there own higher value of He/O, or should we have used He/O= 91 all along, in all of the events? Does the coronal He/O vary from event to event?

The higher value of He/O is consistent with the value 96±35 found in the slow solar wind at solar maximum (von Steiger *et al.* 2000). While the origin of the slow solar wind is not fully resolved, it does have a FIP-effect similar to that of SEPs, which is seen to occur in the plasma near active regions (Brooks, Ugarte-Urra, and Warren 2016). Furthermore, acceleration of large SEP events begins at ~2–3 solar radii (Reames 2009a, b) while the slow solar wind begins in this same region or slightly higher (Abbo *et al.* 2016).

Variations of He/O are seen in the solar wind as functions of time and of solar-wind speed (Collier *et al.* 1996; Bochsler 2007; Rakowsky and Laming 2012) and large variations are seen in H/He with phase in the solar cycle (Kasper *et al.* 2007). Are these abundance variations a property of the solar corona itself, or only related to the formation of the solar wind? Could these, or similar coronal variations, also be measurable in SEPs? To resolve this question for He/O, or any other measurements in SEP events, we must first resolve the source effects as distinct from the (power-law-dependent) transport effects.

In this paper we explore possible variations in the coronal reference He/O abundance by revisiting the SEP events tabulated in Reames (2016a). We use measurements of the relative abundances of the elements He, C, N, O, Ne, Mg, Si, S, Ar, Ca, and Fe and the intervals of atomic number $34 \leq Z \leq 40$, $50 \leq Z \leq 56$, and $Z > 56$, using the *Low-Energy Matrix Telescope* (LEMT; von Rosenvinge *et al.* 1995) on the *Wind* spacecraft. Most abundances were obtained in the interval 3.3–5 MeV amu$^{-1}$, but for Ar and Ca we used the 5–10 MeV amu$^{-1}$ interval which has better resolution (see Reames 2014). Species in the intervals $34 \leq Z \leq 40$, $50 \leq Z \leq 56$, and $Z > 56$ were measured at 3–10 MeV amu$^{-1}$.

Tylka (as quoted in Reames 2013) has described the region of 3-10 MeV amu$^{-1}$ as the "sweet spot" for abundance measurements where the power-law spectra from shock acceleration are least damaged by other effects. Above 10 MeV amu$^{-1}$, spectral breaks are common that depend upon both particle rigidity and the angle between *B* and the shock normal. Fe breaks at much different MeV amu$^{-1}$ than O or He (Tylka et al 2005; Tylka and Lee 2006). Below 1 MeV amu$^{-1}$ in large SEP events, ions are blocked and spectra flattened by transport (Reames and Ng 2010). Many interesting things can be measured at higher and lower energies but 3-10 MeV amu$^{-1}$ may be best for coronal abundances.

## 2. Increasing the Reference He/O from 57 to 91

To what extent can we improve the power-law fit of enhancement *vs. A/Q* by increasing the reference He/O from 57 to 91? Figures 2 through 5 compare the properties of fits using each of these values of reference He/O for four different SEP events.

    The event of 6 November 1997 shown in Figure 2 is typical of many of the 24% of events with source plasma temperatures of 2.5–3.2 MK. These events include reacceleration of residual suprathermal ions from multiple small impulsive SEP events (Reames 2016a, b, c). In panel (e) of the figure it is clear that He lies well above the trend line in the best fits of enhancement *vs. A/Q* in all time periods during the event using the reference He/O abundance of 57. Using reference He/O=91 would drive down these He points into better agreement with C and N below. The effect of the change to He/O=91 is to deepen the minimum in $\chi^2/m$ shown in panel (g) compared with that in panel (d). Changes in the temperatures, panel (f) *vs.* panel (c), are minimal.

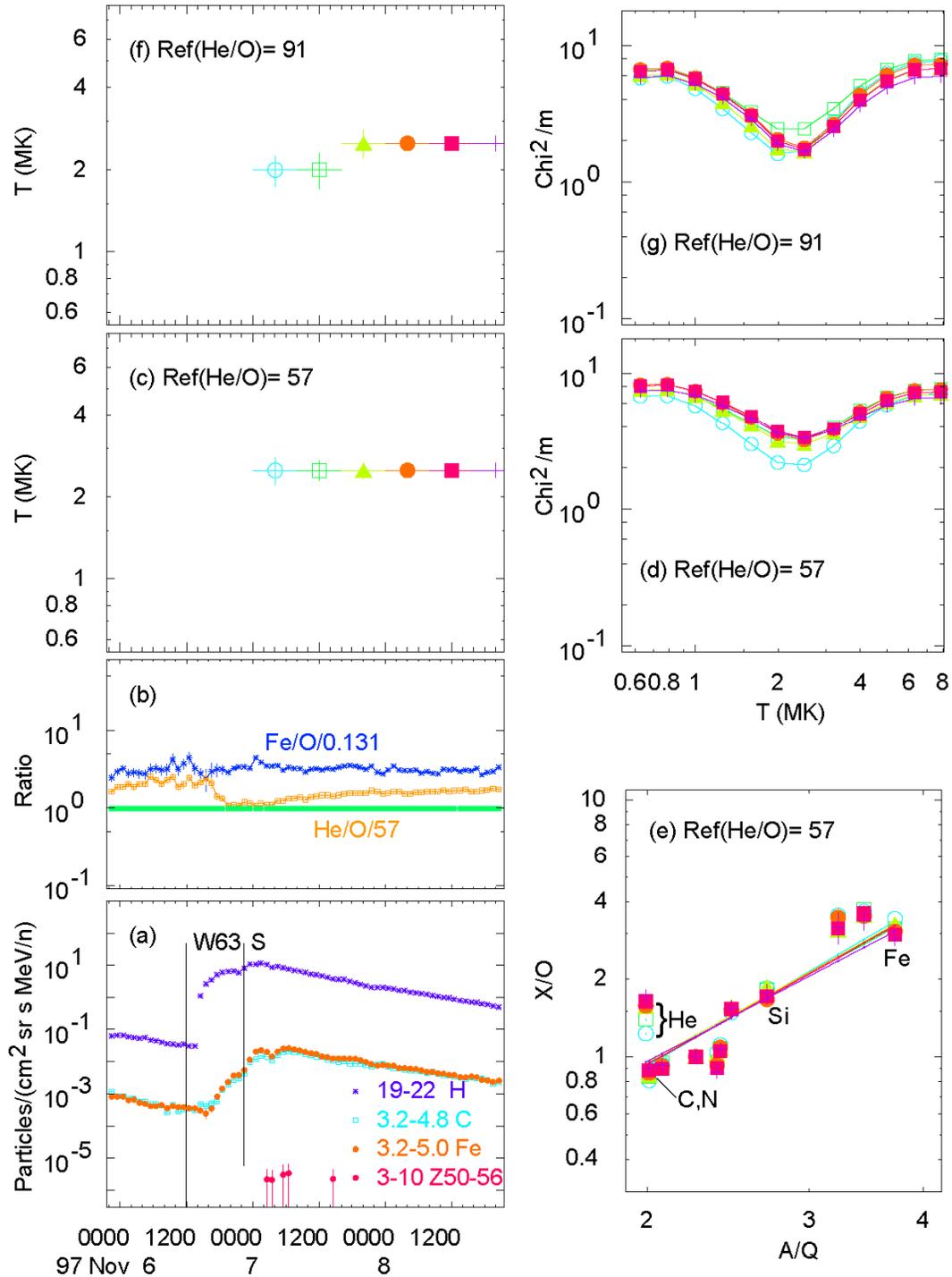

**Figure 2**. For the 6 November 1997 SEP event we show (a) the specified intensities *vs.* time (b) the enhancements in Fe/O and He/O during the event, (c) the best-fit temperatures in color-coded intervals, (d) values of $\chi^2/m$ *vs. T* for each color-coded time interval, and (e) sample best-fit enhancements of X/O, *vs. A/Q*. Panels (c), (d) and (e) are based upon reference He/O=57, while the corresponding panels (f) and (g), based upon reference He/O=91, show substantial improvement in $\chi^2/m$ of the fits.

The event of 20 April 1998 shown in Figure 3 is unique. Using reference He/O of 91 rather than 57 would steepen the slope of the fits in panel (e) and completely remove the ambiguous high temperature and weaken the corresponding upper minimum in panel (d). Note in Figure 1 that at ~8 MK all the elements up to Ne are fully ionized with $A/Q \approx 2$ and the elements up to Ca are highly ionized. Thus whenever all these elements cluster together with small enhancements, $T \sim 8$ MK becomes a good fit.

Actually the event of 20 April 1998 is well known for a complex time-dependent pattern of wave growth and scattering (Tylka, Reames, and Ng 1999; Ng, Reames, and Tylka 1999). Thus it is somewhat surprising that a single solution at 0.8 MK persists, even with changes in He/O.

The Bastille-Day event of 14 July 2000 is shown in Figure 4. Panel (e) of the figure, produced with reference He/O=57, shows that He agrees well with both the ascending and descending power-law fits. Changing the reference He/O to 91 reduces the agreement in both cases, flattens $\chi^2/m$ in panel (g), and increases the errors and ambiguity in $T$ in panel (f), although a value of $T \approx 1.3$ MK does persist. Changing the reference He/O to 91 does not seem appropriate for this event.

The event of 23 August 2005, shown in Figure 5, is another case where increasing the reference He/O seems ill-advised. In panel (e), He already lies well below the power-law trend of the other elements. Increasing the reference He/O to 91 drives He down even farther, and flattens the minimum in $\chi^2/m$ near 1.2 MK in panel (g) so much that the high-temperature solution for stripped ions with no systematic trend from He – Ca dominates.

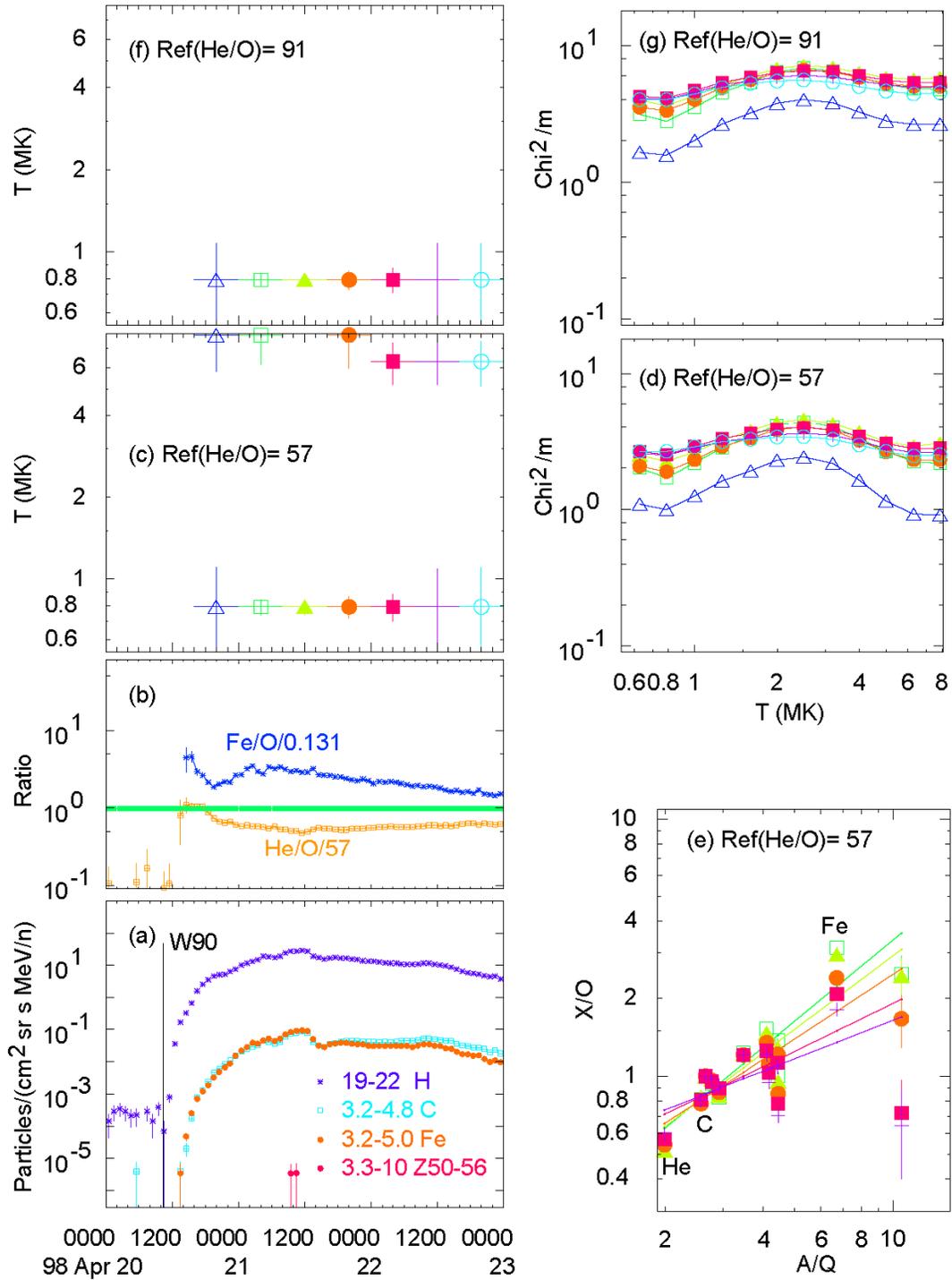

**Figure 3**. Panels (a) through (g) are as described in Figure 2, but for the 20 April 1998 SEP event. The reference He/O=91 reduces the relative importance of the ambiguous hot or stripped ions in $\chi^2/m$.

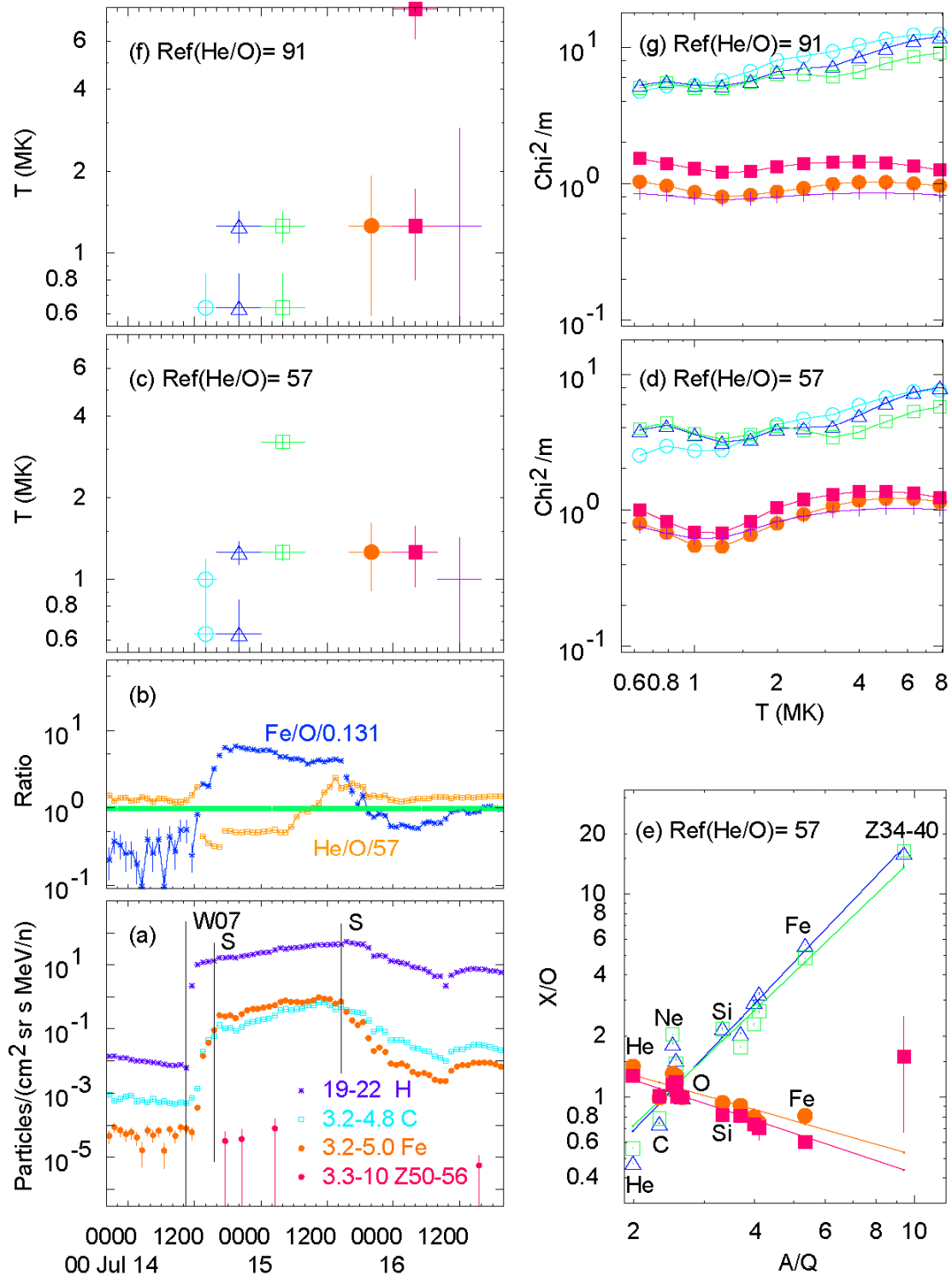

**Figure 4**. Panels (a) through (g) are as described in Figure 2, but for the 14 July 2000 SEP event. Despite some temperature ambiguities, He agrees well with both the ascending and descending fits in panel (e) with reference He/O=57. He/O=91 greatly weakens the $\chi^2/m$ fits in panel (g).

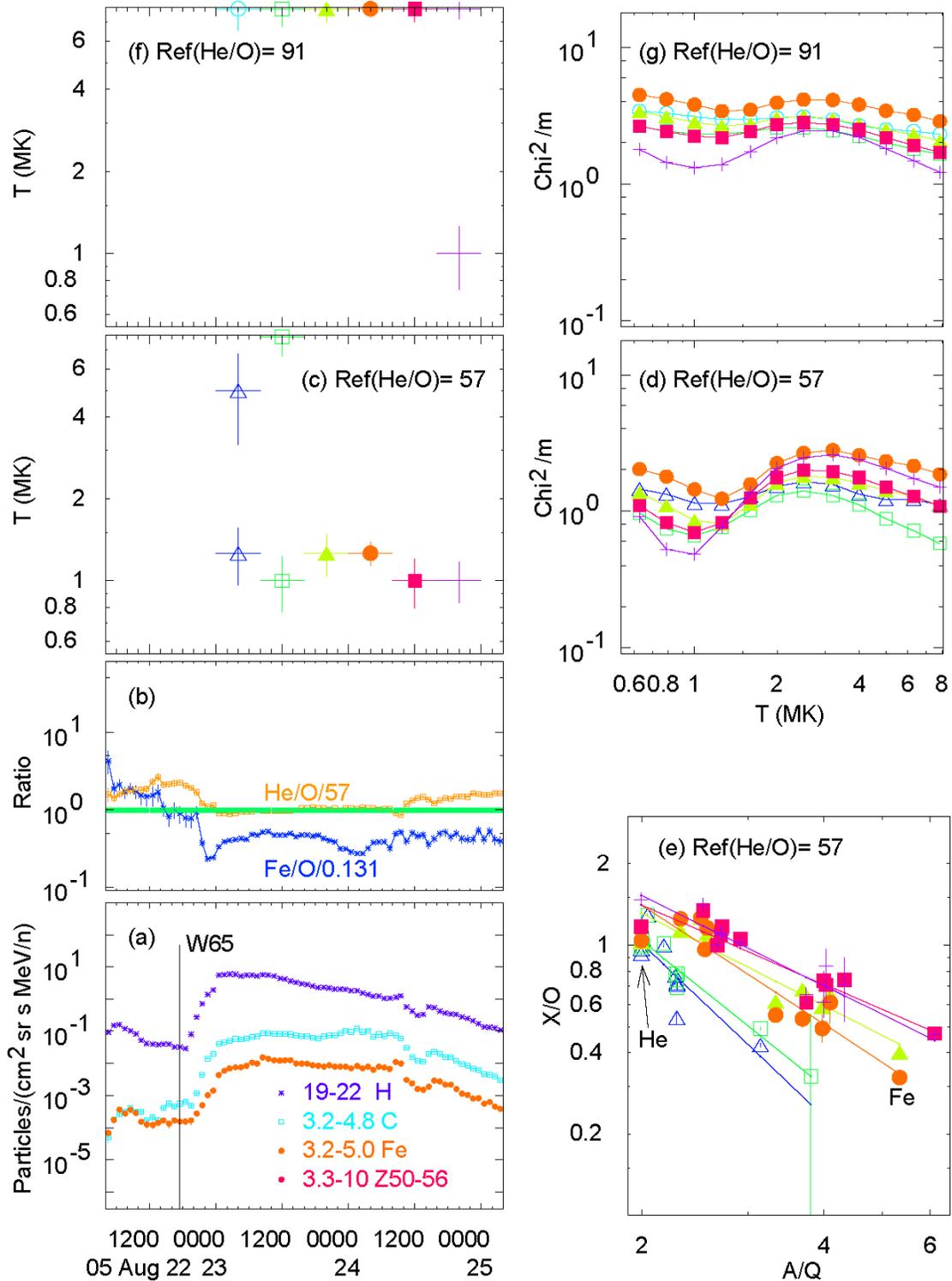

**Figure 5**. Panels (a) through (g) are as described in Figure 2, but for the 23 August 2005 SEP event. Here He is already below all the fit lines with reference He/O=57 in panel (e). Using reference He/O=91 only drives He farther down, destroying the fit near 1 MK.

Using a reference He/O = 91 seem appropriate for events involving active-region plasma, but not for any others. It seems appropriate to ask more generally, does He/O vary from event to event? What would be the best choice for each event?

## 3. The Best Reference He/O Value for Each SEP Event

The study of source plasma temperatures (*e.g.* Reames 2016a) involves nearly 400 8-hr time periods. For each of these periods, we ask what value of reference He/O would bring He onto the best power-law fit line defined by *all* of the elements. The result is shown in Figure 6. Higher values of the coronal He/O seem to be appropriate for SEP events from active regions with $T$= 2–4 MK as found by Reames (2016b) and seen in the event in Figure 2. Events from the temperature region of 1 – 2 MK seem to require a coronal He/O value as low as 40.

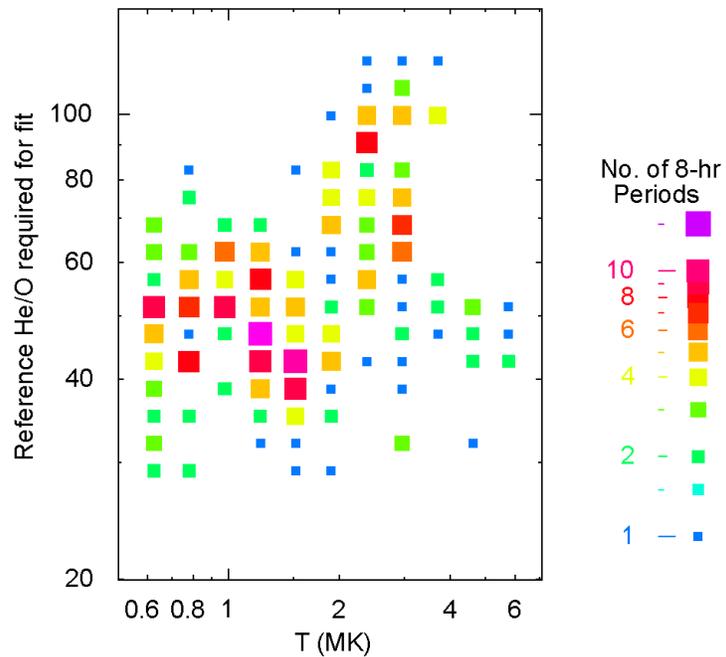

**Figure 6.** A histogram of the reference He/O value that would bring the He enhancement onto the best-fit power-law in each 8-hr period is shown for each temperature interval. Temperatures in the 2.5–3.2 MK active-region interval require higher values of the underlying coronal He/O ratio.

## 4. Decreasing the Reference He/O from 57 to 40

Figures 7 to 9 show a sample of events that benefit by *decreasing* the reference He/O value from 57 to 40.

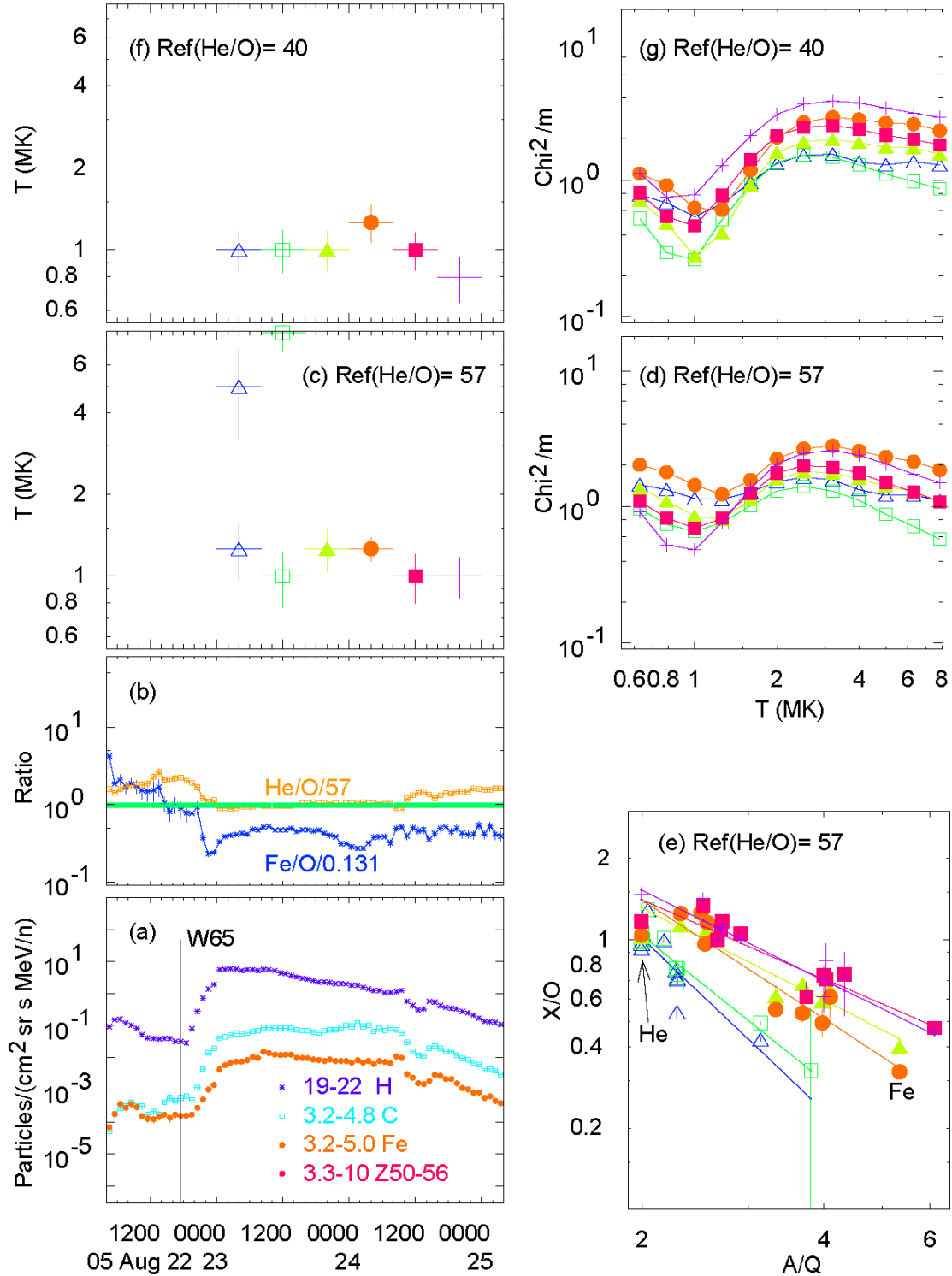

**Figure 7**. Panels (a) through (e) are as described in Figure 2, but for the 23 August 2005 SEP event using He/O = 57 as was shown in Figure 5. However, panels (f) and (g) are now based upon reference He/O = 40. The lower value of reference He/O moves He up toward the fit lines in (e), deepens the minimum $\chi^2/m$ near 1 MK in panel (g) and removes the early temperature ambiguity.

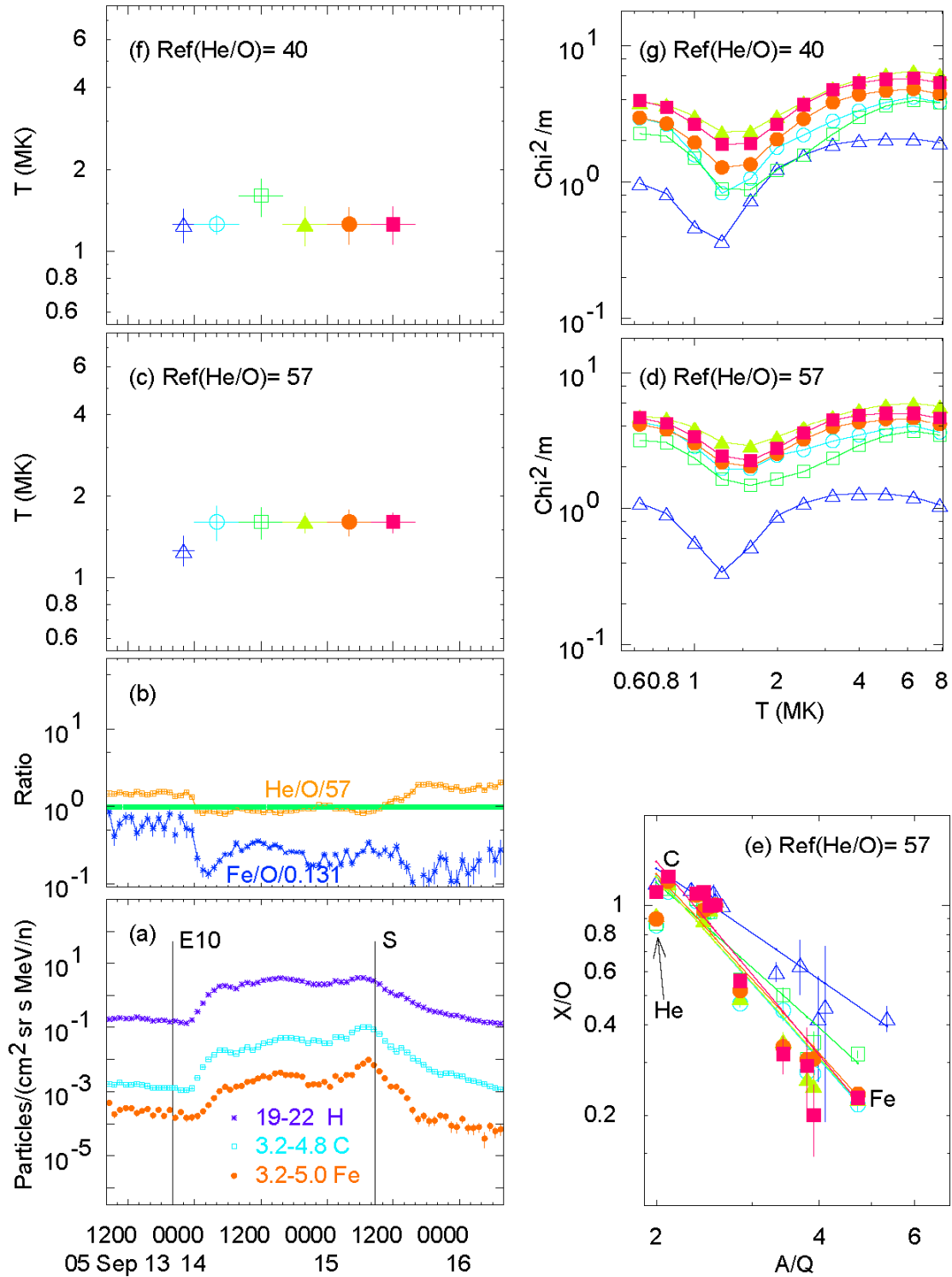

**Figure 8**. Panels (a) through (e) are as described in Figure 2, but for the 13 September 2005 SEP event using reference He/O = 57. Panels (f) and (g) are based upon He/O=40. He falls below the fits in panel (e) and is raised, improving the fit, by using the smaller reference value.

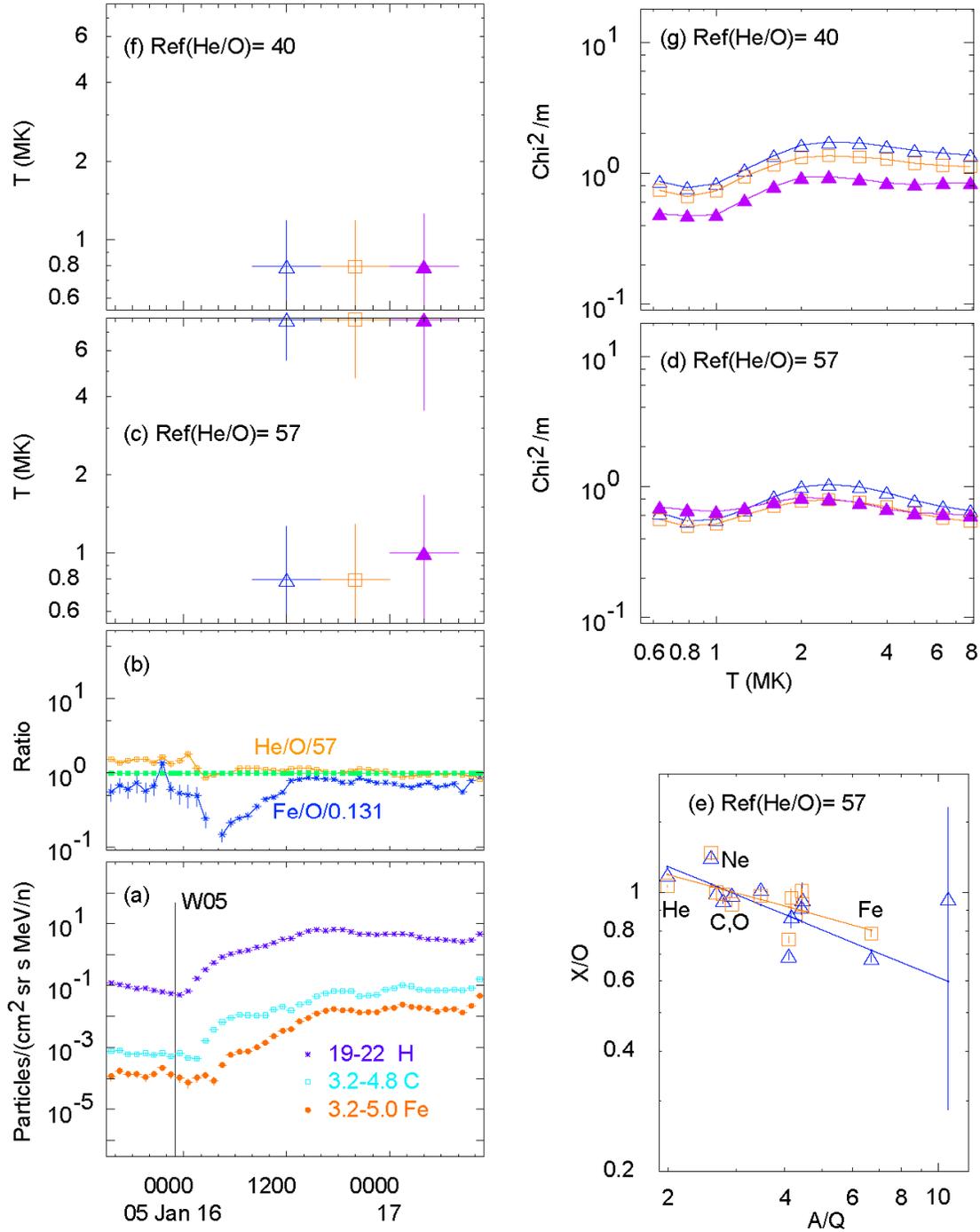

**Figure 9**. Panels (a) through (e) are as described in Figure 2, but for the 16 January 2005 SEP event using reference He/O = 57. Panels (f) and (g) are based upon He/O=40. The slopes of the fits in panel (e) are small; lowering the reference He/O to 40 is enough to improve the fits, lower the minimum $\chi^2/m$ near 0.8 MK in panel (g) and completely remove the temperature ambiguity.

In each case, lowering the reference He/O to 40 shifts He *toward* the fit line, improves the minimum $\chi^2/m$ of the best-fit power law and removes any temperature ambiguity that exists. Reames (2016a) found a gathering of SEP events, all with $T$< 2 MK, which had evidence of "stripped" ions. While an unenhanced cluster of ions from He through Ne can be produced by adding hot or stripped population of ions, it can also be artificially produced by using the wrong value of the reference He/O which can shift He away from its proper fit line, especially when the slope of the fit line is small. It seems probable that the evidence of "stripped" ions was spurious and was never real.

We have found that the underlying "coronal" abundance of He/O varies with time and place.

## 5. Variations in the Reference He/O

In this section we compare SEP events with extreme values of the coronal He/O ratio and explore the relative constancy of that ratio during the events. Figure 10 compares the extremely low value of He/O needed to fit the power-law for the low-temperature event of 13 September 2004 ($T$ = 1.4±0.2 MK) with a typical Fe-rich, 2≤ $T$ ≤4 MK, event, that of 6 July 2012 ($T$ = 2.5±0.3 MK). The reference value of He/O used to obtain the best-fit power law in each time period is listed in the upper two panels for each SEP event. The upper panel then shows the value of the coronal He/O that would be required to put the observed He/O on that best-fit line in that time period. The high value of He/O for the event of 6 July seems to continue into the onset of the event on July 8, indicating that both events have access to the impulsive suprathermal ions in the same active region.

Figure 11 compares the low-temperature event of 30 September 2013 ($T$ = 1.3±0.3 MK) with the well-known Fe-rich event of 4 November 1997 ($T$ = 2.7±0.5 MK). For the 30 September event from W33, panel (b) shows Fe/O that is enhanced early and declines to a deeply depressed value later, typical of many events from western sources that have minimal access to impulsive suprathermal ions. Note that the 30 September 2013 event and the 4 November 1997 event both have sources at W33. The seed population for the latter must contain impulsive suprathermal ions from a solar active region, while the seed population for the former consists only of coronal plasma at 1.3 MK.

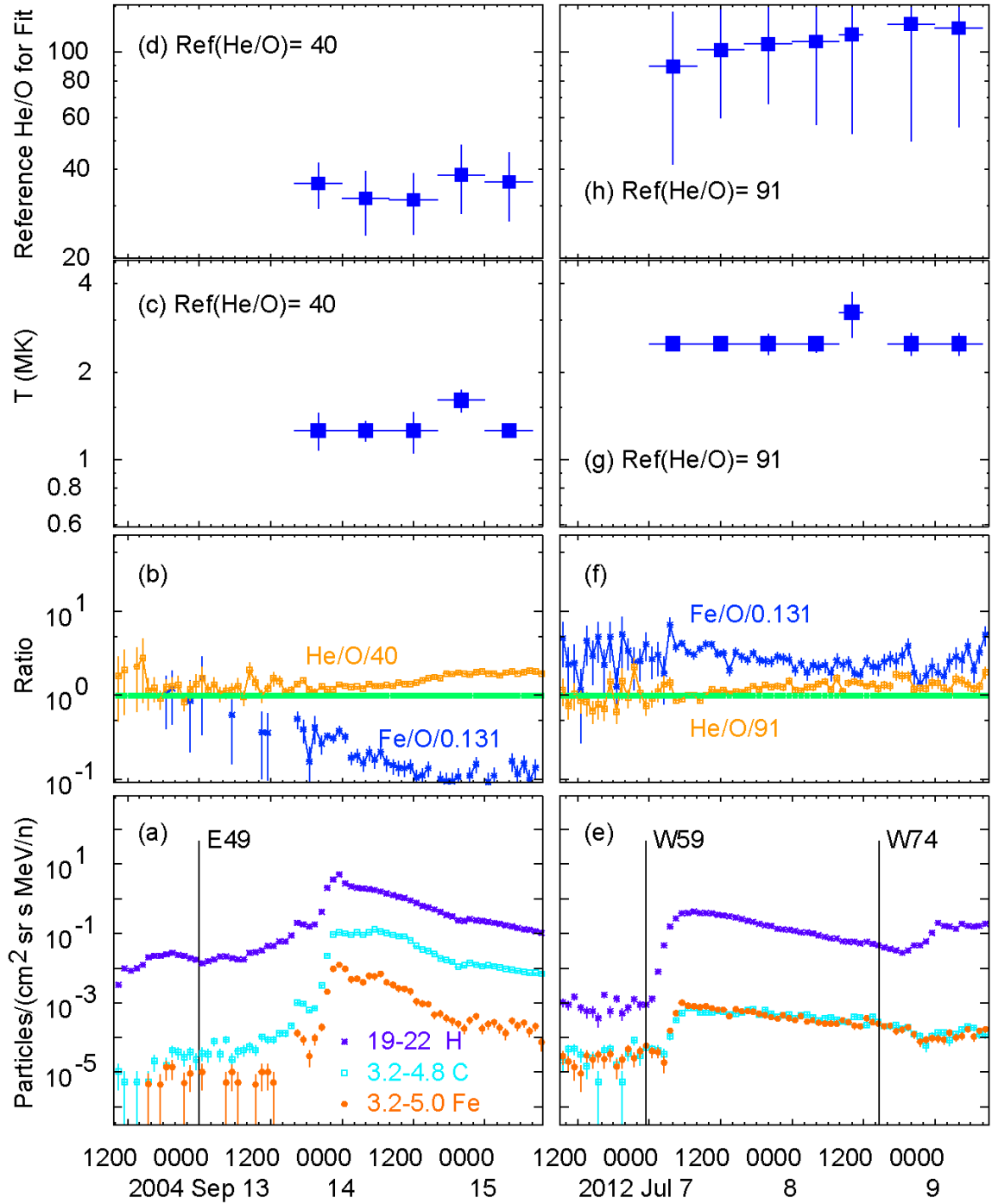

**Figure 10** compares two SEP events with extreme values of the coronal He/O ratio required to agree with the power-law fit of enhancement *vs. A/Q*. For the event of 6 July 2012, panel (a) shows selected intensities, (b) shows the relative values of Fe/O and He/O, (c) shows the derived source temperature, and (d) shows the derived coronal He/O, all *vs.* time. Panels (e) – (h) show corresponding parameters for the Fe-rich 6 July 2012 SEP event. Values of derived coronal He/O are reasonably consistent with those originally assumed to fit each event.

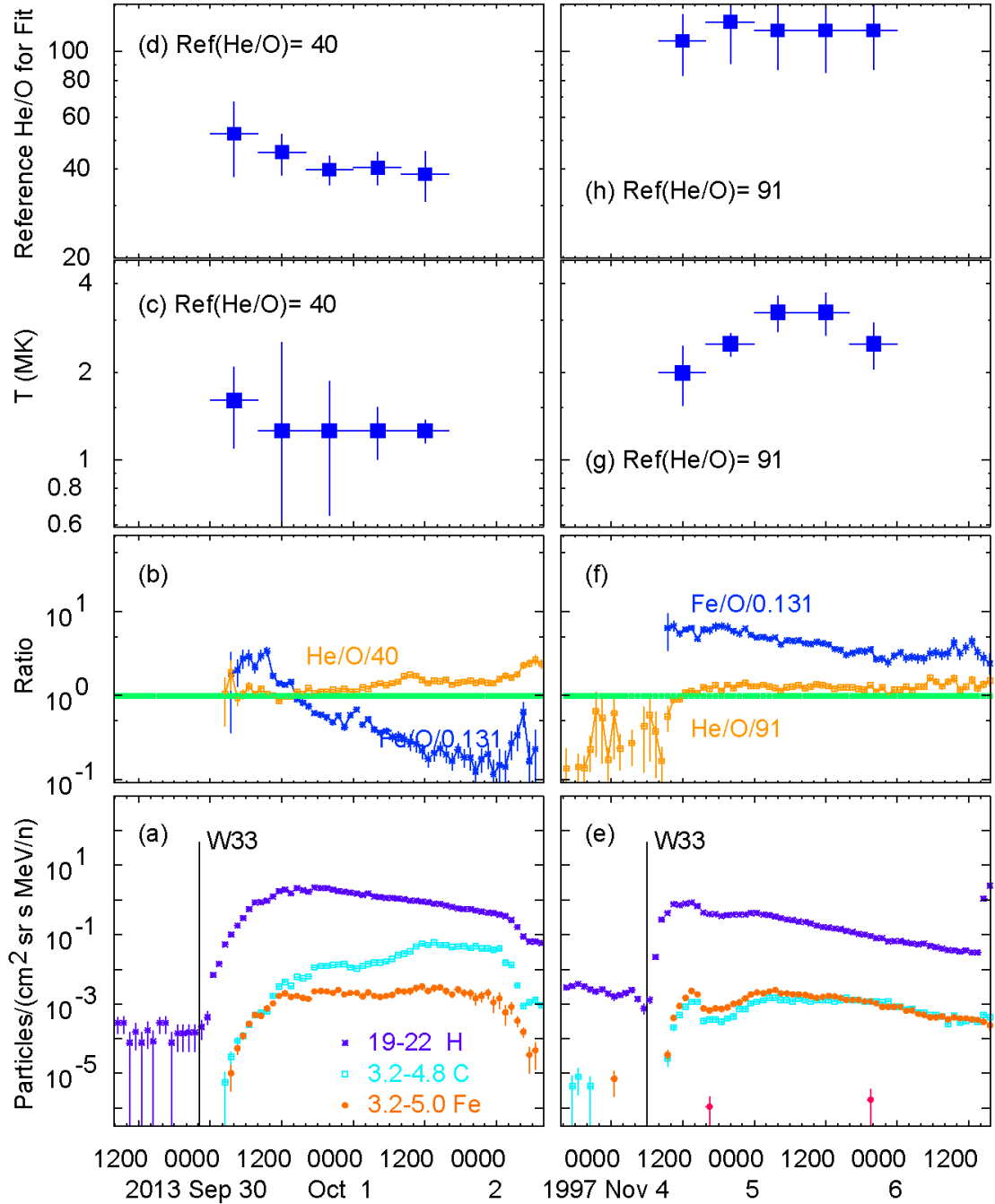

**Figure 11** compares two SEP events with extreme values of the coronal He/O ratio required to agree with the power-law fit of enhancement *vs. A/Q* as in Figure 11. Panels (a)–(d) are for the 30 September 2013 SEP event and panels (e)–(h) for the Fe-rich SEP event of 4 November 1997. Note that both events are from sources at W33 on the Sun, yet they have different access to impulsive suprathermal ions in their seed population.

In Figure 12 we summarize the estimated coronal values of He/O for each event as a function of time with the source temperature indicated.

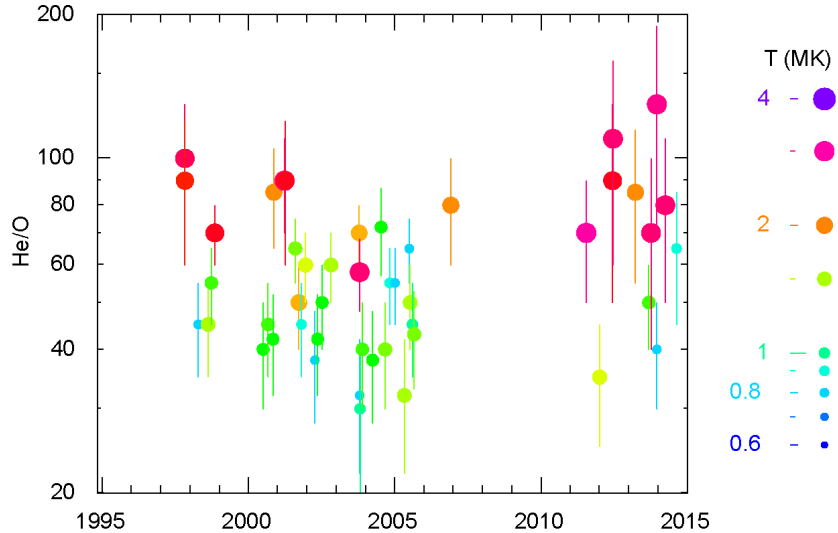

**Figure 12.** The estimated coronal value of He/O is shown *vs.* time in gradual SEP events, with the source plasma temperature indicated as the size and color of the symbol.

The SEP events with $T$ = 2–4 MK have values of He/O that are systematically higher than events with $T$ < 2 MK. These hotter events also tend to be more prevalent early in each solar cycle, perhaps because CMEs are more likely to occur near active regions early in each cycle.

It is difficult to see any trend in He/O for SEP events with $T$ < 2 MK. He/H in the slow solar wind decreases by a large factor between 2003 and 2005, also depending strongly upon flow speed and the solar cycle (Kasper *et al.* 2007). He/O variations are also seen in the solar wind (Collier *et al.* 1996; Bochsler 2007; Rakowsky and Laming 2012) as a function of wind speed but they lack the strong solar-cycle dependence of He/H, as we also see for He/O in SEPs in Figure 12.

In Alfvén-wave models of the FIP effect (Laming 2009; Rakowsky and Laming 2012; Laming 2015), He, with the highest ionization potential, stays neutral longest and is therefore preferentially depleted, relative to other elements as the ions are driven up into the corona. In any case, the mechanisms for depleting or varying He/O occur at the base of the corona and are likely to affect both SEPs and the solar wind.

## 6. Discussion

In this study, we have found that the "coronal" He/O ratio, as sampled by SEP events, is not constant, but can vary widely from event to event. Fe-rich SEP events, with source plasma temperatures of 2–4 MK, associated with re-acceleration of residual suprathermal ions generated by almost continuous emission from multiple small impulsive SEP events in jets from active regions, have He/O ≈ 90, comparable with the average ratio seen in the slow solar wind (*e.g.* He/O= 90±30; Bochsler 2009). Other SEP events, with $T$ < 2 MK and 40 ≤ He/O ≤ 60, seem to come from outside active regions. These events with $T$ < 2 MK lack any significant contribution of impulsive suprathermal ions that might be expected above active regions; they show only the presence of ambient coronal material in the seed population accelerated by the shock.

Reames (2016a) showed two source-temperature regimes for gradual SEP events, but assumed that all abundance variations, including He/O, came from power-law transport effects, despite a questionable fit for He/O. Here, we have found that there are also *coronal* variations in He/O, and that they are *correlated* with the temperature differences of the two regimes.

Large Fe-rich gradual SEP events have been associated with active regions (Ko *et al.* 2013) and regions of persistent emission of small impulsive SEP events and with individual sources too small to identify have been seen often (Desai *et al.* 2003; Bučík *et al.* 2014, 2015; Chen *et al.* 2015). The temperature of the 2–4 MK plasma in the large Fe-rich SEP events originally comes from the source of the small impulsive SEPs (Reames, Cliver, and Kahler 2014a, 2014b) which were accelerated relatively low in the corona (often < 1.5 $R_S$; diFabio *et al.* 2008) in or near active regions.

Regions with a large FIP-bias are known to be associated with solar active regions (Brooks, Ugarte-Urra, and Warren 2016). Plasma temperatures above these regions are ~1.4–1.8 MK (*e.g.* Brooks and Warren 2012; but see also Xie *et al.* 2017). The slow solar wind with its high FIP bias is likely to receive contributions from these regions. The altitude associated with the start of the slow solar wind is ~2 $R_S$ (Abbo *et al.* 2016). The altitude for the onset of large SEP events is also ~2 $R_S$ (Reames 2009a, 2009b). Thus the shock waves that accelerate the SEPs probe the same region that is the origin of the slow solar wind. Also, a single CME-driven shock wave is broad enough to sample plasma from different regions of the corona.

It is difficult to connect the source plasma temperatures derived from the SEPs directly with the solar wind since the solar wind is not isothermal. However, if we assume that those temperatures are closely related to the "freezing-in" temperatures for C and O, we can determine theoretical ratios such as $O^{+7}/O^{+6}$ from the SEP temperatures. At a temperature of 2 MK, $O^{+7}/O^{+6} \approx 1$ (Mazzotta *et al.* 1998), typical of the slow solar wind produced by sector reversal, pseudo-streamers, and magnetic clouds (Xu and Borovsky 2015; see also Fu *et al.* 2017); at $T <$ 1 MK, $O^{+7}/O^{+6} <$ 0.01, typical of the fast wind from coronal holes. However, accelerated material from previous CME ejecta must also contribute to some events. From this rough comparison, most of the SEP events fall into the source temperature region that corresponds to the slow solar wind.

Many cooler regions sampled by SEPs show differences in temperature and in He/O, but no evidence of a difference in the level of FIP-bias. Shocks *do* propagate into the fast solar wind, but there is no evidence there of SEPs with lower FIP-bias (Kahler and Reames 2003, Kahler, Tylka, and Reames 2009). Average abundances of SEPs observed in fast wind are statistically the same as those in the slow wind. To what extent can SEP abundances be transported laterally across the shock surface? Reames (2017b) found one event that showed the same unique non-power-law abundance pattern across 222° of solar longitude. This event, on 23 January 2012, had flat or increasing abundance enhancements from He to Mg then decreasing abundances up to Fe. Such a pattern could be produced by a low-energy *A/Q*-dependent spectral break preferentially suppressing heavier elements like Fe (Tylka and Lee 2006, Reames and Ng 2010), but it is surprising to see the same pattern so extensively across the shock. This apparent lateral transport seems to conflict with evidence in the event of 31 August 2012, discussed below, that shows an SEP source temperature of 3.2 MK at one spacecraft and 1.6 MK at another, showing no mixing over 116° of longitude in this case for at least two days.

It is important to realize that our association of the Fe-rich source plasma at $T =$ 2–4 MK with reacceleration of impulsive suprathermal ions from jets is based upon a long history of studying the uniqueness of $^3$He-rich SEP events. Shock acceleration cannot significantly enhance $^3$He/$^4$He. It is extremely difficult to enhance $^3$He/$^4$He by factors of ~1000 without a resonance mechanism like that of Temerin and Roth (1992, see also Drake *et al.* 2009) where the waves that enhance $^3$He are generated by the same impulsive electron beams that produce type III radio bursts. The $^3$He-rich – type III associations begins with Reames and Stone (1986). Impulsive SEP events generally lack shocks and type II radio bursts. The story of $^3$He in gradual events begins with Mason, Mazur, and Dwyer (1999) and reacceleration of impulsive suprathermal ions is the basis of Tylka *et al.* (2005) and Tylka and Lee (2006). There is a long, 40-year, history of association of $^3$He-rich and Fe-rich events that begins with Mason *et al.* (1986) and Reames, Meyer, and von Rosenvinge (1994) and continues through generations of instrument development to Reames, Cliver, and Kahler (2014a and b). The association of $^3$He-rich events with *narrow, slow* CMEs and the open fields of jets begins with Kahler, Reames, and Sheeley (2001, see also Reames 2002). The sources and properties of $^3$He-rich, Fe-rich events have been reviewed by Mason (2007), by Reames, Cliver, and Kahler (2014a and b) and by Reames (2017a). Recent papers tend to define "impulsive" events using Fe/O rather than $^3$He/$^4$He because the former better characterizes the whole event while the latter varies strongly with energy (Mason 2007). The necessity for multiple small jets to average the abundance variations in gradual events was shown by Reames (2016b) and their existence by Desai *et al.* (2003), Bučík *et al.* (2014, 2015), and Chen *et al.* (2015). After a legacy of many solar-cycles, the unique character of the $^3$He-rich, Fe-rich material now provides a powerful signature of the well-documented physics of its origin.

While the origin of the Fe-rich SEP events with $T$ = 2–4 MK is relatively clear, the origin of the SEP events with $T$ < 2 MK is less clear. Multi-spacecraft observations (Reames 2017b) show one event on 31 August 2012 where a spacecraft that is well-connected to the source longitude sees SEPs from $T \approx 3.2$ MK while a poorly-connected spacecraft, 116$^o$ to the west, sees those from $T \approx 1.6$ MK. This suggests that the shock accelerates hotter material from residual impulsive suprathermal ions from jets above the active region and cooler material elsewhere, as shown in Figure 13. Contributions from many small jets are needed to average out the large fluctuations seen in individual impulsive SEP events but not seen in gradual SEP events (Reames 2016b, 2016c). The impulsive suprathermal ions from previous jets are preferentially accelerated as SEPs by shock waves, and it is possible that the bulk plasma ejected in jets contributes to the slow solar wind, although the underlying abundances in each case need not be identical.

Recently, Xie *et al.* (2017) have measured temperatures of 50 coronal loops in 42 active regions. They found 39 loops with $T$ = 1–2 MK and 11 loops with $T$ < 1 MK, not unlike the distribution we find from SEP events. However, we should reemphasize that we are defining "active regions" with $T$ = 2–4 MK by the presence of SEPs with unique ($^3$He-rich, Fe-rich, Z>50-rich) abundances reaccelerated from the impulsive seed popula-tion strongly associated with flares or jets from solar active regions (Kahler, Reames, and Sheeley 2001; Reames, Cliver, and Kahler, 2014a, 2014b, 2015).

It is possible to treat abundances of other elements as we have treated He, *i.e.* to ask what coronal abundance would bring that element to the power-law fit line defined by the other elements. However, for other elements the variations are smaller and the errors larger so it is difficult to find a value for each event. For C/O, mean values show essentially no temperature difference: derived coronal C/O = 0.392±0.007 averaged over SEP events with $T$ = 2–4 MK and C/O = 0.398±0.006 for $T$ < 2 MK. However, this ratio for SEPs is *never* seen to exceed 0.50, nor approach the mean value of C/O = 0.68±0.07 in the solar wind (Bochsler 2009). C/O shows the largest difference between SEPs and the slow solar wind, a difference that remains unexplained, but a more recent measurement in the bulk solar wind does give C/O=0.53±0.06 (Heber *et al.* 2013).

Finally, in Figure 13 we compare the average SEP abundances with the slow (interstream) solar wind (Bochsler 2009), both relative to photospheric abundances (Asplund *et al.* as a function of FIP. We use the SEP reference abundances corrected with He/O=91. Relative to the SEPs, the SW abundances of P and C are high, while N is low. Other abundances are in statistical agreement. However, the more recent measurements in the bulk solar wind give C/O=0.53±0.06 and N/O=0.095±0.007 (Heber *et al.* 2013), in somewhat better agreement with SEPs.

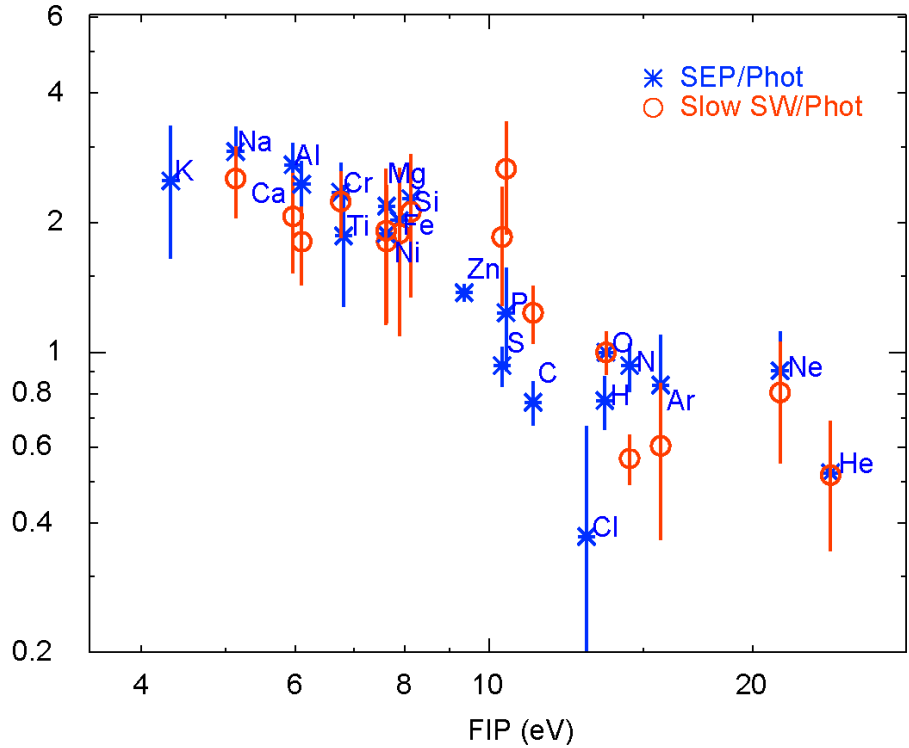

**Figure 13.** Abundances of SEPs and the slow (interstream) solar wind (Bochsler 2009), relative to the photosphere (Aslpund *et al.* 2009), are shown vs. FIP. Compared with SEPs, SW abundances of C and P are high, and N is low (see text). Based upon Fe/O alone, the FIP levels of SEPs and slow SW are nearly identical.

## 7. Summary

The major finding of this study is that the coronal value of He/O sampled by SEP events can vary from event to event, *i.e.* from place to place or time to time in the corona. For the events with 2–4 MK source plasma temperatures, dominated by impulsive suprathermal ions that have been reaccelerated by the shock wave, He/O ≈ 90. In fact, both the higher temperature and He/O are determined at the source of the impulsive material, primarily in solar jets. For the SEP events with $T < 2$ MK, where shock waves accelerate ambient coronal plasma, we find $40 \leq$ He/O $\leq 60$, uncorrelated with $T$. In any case, the mechanism producing the He/O variations operates near the base of the corona (~1.1 $R_S$) and is thus likely to produce the same abundances in SEPs (at ~2 – 3 $R_S$) and in the solar wind.

  A previous suggestion that apparently-high temperatures, with clustered enhancements, might represent a component of stripped ions now seems invalid. This pat-tern was probably produced by inappropriate values of He/O that disrupted the trend of the power law dependence on *A/Q*.

**Acknowledgments:** The author thanks Steve Kahler for many helpful discussions of subjects included in this manuscript.

## Disclosure of Potential Conflicts of Interest

The authors declare they have no conflicts of interest.